\newif\ifEXTENDEDVERSION
\EXTENDEDVERSIONtrue
\EXTENDEDVERSIONfalse

\documentclass[a4paper,UKenglish,cleveref, autoref,thm-restate]{lipics-v2021}

\hideLIPIcs
\usepackage{booktabs}
\usepackage{rotating} 
\usepackage{listings} 
\usepackage{graphicx}  
\usepackage{wrapfig}   
\usepackage{booktabs}  
\usepackage{cleveref}  
\usepackage{xspace}

\newcommand{\mysubsection}[1]{\subsection{#1}}

\newcommand{\eg}{\textit{e.g.},\xspace}

\title{Software Supply Chains are Dead: Use-Case-Oriented Regeneration}

\titlerunning{Software Supply Chains are Dead: Use-Case-Oriented Regeneration}

\author{Tanmay Singla}
{Purdue University}
{singlat@purdue.edu}
{}
{}

\author{James C. Davis}
{Purdue University}
{davisjam@purdue.edu}
{}
{}

\authorrunning{Singla \& Davis}

\Copyright{Anonymous Author(s)}



\category{}

\relatedversion{}

\acknowledgements{We acknowledge support from the US National Science Foundation under award \#2541917.}

\nolinenumbers 

\EventEditors{John Q. Open and Joan R. Access}
\EventNoEds{2}
\EventLongTitle{42nd Conference on Very Important Topics (CVIT 2016)}
\EventShortTitle{CVIT 2016}
\EventAcronym{CVIT}
\EventYear{2016}
\EventDate{December 24--27, 2016}
\EventLocation{Little Whinging, United Kingdom}
\EventLogo{}
\SeriesVolume{42}
\ArticleNo{23}

\begin{document}

\maketitle

\begin{abstract}
Modern software development relies on an increasingly doubtful premise:
  that the up-front implementation savings from adopting a dependency outweighs the maintenance costs.
Two changes are reshaping the build-\textit{vs.}-reuse calculus:
  software supply chain attacks have raised the cost of external reliance,
  while generative AI has lowered the cost of local implementation.
We envision \emph{use-case-oriented regeneration} as a new software sourcing paradigm that shifts the supply chain from external trust to local verification.
We evaluate an agentic workflow that synthesizes only the specific slice of dependency functionality that a repository exercises. 
Our measurements across 180 repository--dependency pairs suggest that this approach is feasible:
  the replacements preserve 99.8\% of repository-observed behavior across baseline validation checks and reduce the exported API surface by 93\%.
Software sourcing may evolve toward verifiable repository-specific code synthesis, especially when the required functionality is narrow, stable, and well tested.
\end{abstract}

\section{Introduction}
\label{sec:introduction}

For over two decades, software reuse has been increasingly organized around package-level dependency adoption.
When software engineers need functionality, the default posture is to depend on upstream packages, even when the repository uses only a small part of that package's behavior.
This artifact-centric model helped make reuse inexpensive and scalable~\cite{wittern2016look,zimmermann2019npm}, but it also creates a persistent granularity mismatch: downstream repositories adopt, validate, and maintain entire upstream artifacts to obtain narrow pieces of functionality.
That mismatch has become increasingly costly as repositories accumulate transitive dependency graphs and inherit upstream vulnerabilities, maintenance decisions, compatibility constraints, and ecosystem-level risks~\cite{paschal_ztd, chinenye_taylor_scored, kelechi_usenix25}.
At the same time, generative AI and coding agents are lowering implementation costs as they become more capable of inspecting repositories, executing validation tools, and iteratively refining code~\cite{yang2024sweagent, wang2024openhands, jimenez2023swe}.
Together, these changes reopen the build-\textit{vs.}-reuse decision: for narrow, stable uses of a dependency, relying on the full upstream artifact may no longer be the best choice.

We envision \emph{use-case-oriented regeneration} as a new software sourcing paradigm that shifts dependency management from external trust to local verification.
Instead of taking on a full dependency, an engineer can now use an agentic workflow to synthesize only the specific functionality that is actually needed as a local, repository-owned implementation.
Building on prior regeneration work, we shift the unit of analysis from the upstream package to the \emph{repository--dependency pair}. 
We report on the feasibility of this regenerative paradigm, evaluating regeneration across 180 repository--dependency pairs spanning nine diverse JavaScript/TypeScript dependencies.
Our results show that agentic workflows can successfully replace full dependencies with narrow, local implementations while preserving 99.8\% of repository-observed behavior across baseline validation checks.
When successful, these generated replacements shed the unused code, transitive dependencies, and maintenance obligations that accompany artifact-centric reuse. 

The shift to regeneration creates new responsibilities --- it moves the trust problem from evaluating a persistent upstream maintainer~\cite{kalu2026arms} to validating the correctness and maintainability of locally synthesized code.
This paradigm shift introduces open challenges regarding validation coverage, semantic equivalence, and how developers can establish trust in regenerated code. 
In sum, this paper makes the following contributions:
\begin{itemize}
    \item We propose \emph{use-case-oriented regeneration} as a repository-centric sourcing strategy that competes with conventional full-artifact reuse. We demonstrate its feasibility through an evaluation of 180 repository--dependency pairs, observing behavioral preservation with substantial footprint reduction.
    \item We outline a research agenda for synthesize-and-verify reuse, focusing on appropriateness, behavioral validation, provenance, and refresh practices for generated replacements.
\end{itemize}

\section{Background and Motivation}
\label{sec:background}

\subsection{Dependency-Mediated Reuse and the Build-\textit{vs.}-Reuse Tradeoff}
\label{subsec:dependency-reuse}

Package ecosystems have made reuse cheap by shifting functionality into upstream artifacts.
Instead of building every component from scratch, developers assemble applications from packages distributed through ecosystems such as npm, PyPI, Maven, and Cargo~\cite{wittern2016look,zimmermann2019npm}.
Industry reports estimate that open-source components underpin most modern software --- Black Duck reports that 96\% of audited commercial codebases contain open-source components~\cite{linuxfoundation2022censusii,synopsys2024ossra}, and the Linux Foundation suggests a substantial footprint, with OSS comprising 70--90\% of modern software.
Third-party package dependencies are a primary mechanism by which software functionality is obtained, shared, and maintained.

Package reuse imports obligations beyond the functionality used.
A downstream repository inherits not only the immediate functionality of a dependency, but also its transitive dependency graph, release practices, maintenance state, compatibility constraints, and vulnerabilities~\cite{ohm2020backstabber,enck2025research,Jayasuriya2024Syntatic}.
Dependencies evolve independently of the downstream projects that rely on them.
A package update may fix a vulnerability, introduce a breaking change, or require downstream adaptation, while failing to update may leave a project exposed to known defects~\cite{he2023automating,jafari2023dependencyupdatestrategiespackage}.
Package ecosystems have also become attack surfaces: Sonatype reports more than 1.23 million known malicious open-source packages across major registries, including more than 454,000 newly identified during 2025~\cite{sonatype2026soss}.
Dependency reuse is therefore a sourcing, maintenance, and assurance commitment.

The insight underlying this paper is that \textit{dependencies are often adopted at a coarser granularity than they are used}.
Package ecosystems distribute reusable artifacts as general-purpose packages, but a downstream repository may exercise only a small subset of a package's API or behavior.
The repository must still adopt, validate, and maintain the full artifact to obtain that subset.
This mismatch motivates us to move beyond reactive reduction techniques and reconsider whether continued runtime reliance on full-package artifacts remains necessary for narrow dependency use cases~\cite{weeraddana2024dependency,abdalkareem2020impact,liu2025commonjsbloat}.

This granularity mismatch changes the build-\textit{vs.}-reuse tradeoff.
Software sourcing is traditionally framed as a choice between implementing functionality locally or obtaining it from an external component~\cite{boehm1984software}.
Building locally gives the downstream project control over implementation, integration, review, and evolution, but requires design, implementation, and testing effort.
Reusing an external package reduces the initial cost of acquiring functionality, but introduces costs related to integration, assurance, updates, and dependency management~\cite{vargas2020dep}.
The tradeoff is therefore not only whether code exists, but who owns the implementation, how it is validated, and how it evolves.

The software engineering community is biased toward reuse; even the US Department of Defense approves of open-source software.
We believe generative systems demand that we revisit that tradeoff.
Large language models can inspect repositories, execute validation tools, and iteratively refine code, reducing the acquisition cost of local implementation by generating code for the surrounding repository context~\cite{wang2024openhands,yang2024sweagent,marron2026agenticinfusedsoftwareecosystem}.
Reuse becomes less attractive when one's repository uses a narrow and stable slice of the package, or when the package introduces transitive dependencies and security exposure that exceed the value of the functionality it provides~\cite{ponta2021used}.
By lowering the cost of code creation, generative systems expose another sourcing option:
regenerating the verified slice of dependency functionality that the repository needs.

\subsection{Motivating Example: A Case Study in Generative Sourcing}
\label{sec:background:davis}

For a concrete illustration of this concept, let us share a recent experience. 
We were developing an automated accessibility tool that needed to manipulate mathematics in Microsoft Office documents~\cite{davis2026cheap}. The system had MathML and wanted to render it as OMML for native support in Microsoft Office. Under the traditional sourcing model, we initially treated this as a dependency-selection problem. MathML-to-OMML conversion is not a trivial formatting task: it requires handling structured mathematical constructs, including operators, fractions, matrices, fences, scripts, annotations, and other element types. As good engineers, we therefore assessed existing implementations, including open-source converters, document-editor subprocesses, and commercial options.

Some candidates were plausible under the conventional reuse calculus:
  credible provenance, acceptable licensing, or good pilot behavior on our fixtures.
However, that calculus assumes that the costs of taking on the dependency are lower than the cost of developing it ourselves.
In our case, adding an external converter would have entailed more elaborate component vetting, licensing concerns, and bloating our Google Cloud images with the JVM.\footnote{Image bloat would reduce our service's peak scalability under our current Cloud tenant memory quota.}
We therefore considered an alternative: using an AI coding agent to implement the converter locally.
We asked Claude about the complexity of MathML, which it characterized as involving roughly thirty relevant node types; and about the feasibility of a straightforward tree-transducer approach (whether each type could convert cleanly to OMML without reasoning over an intermediate representation).
Receiving positive answers to each, we asked Claude to construct a MathML-to-OMML library of our own, backed by fixtures and validation against Office's OpenXML SDK.
After Claude spent an hour on it, the resulting library worked on the first try with our validation fixtures.
Claude's library has 1.4\ KLOC, accompanied by 1.5\ KLOC of unit and property-based tests.
On inspection, the library code is of good quality --- structured and concise.

This anecdote is not evidence that every converter should be regenerated.
Rather, it illustrates the changed decision problem for package selection when the needed behavior is bounded and testable.
\textit{Is relying on existing packages still preferable once agent-assisted local synthesis makes a small, auditable, product-specific implementation cheap enough to create?}

\subsection{Dependency Reduction and Internalization Strategies}
\label{subsec:reduction-strategies}

Several existing strategies already attempt to mitigate the cost or exposure associated with overly broad dependencies. Debloating, for instance, removes unnecessary code from software artifacts or deployed systems while attempting to preserve required behavior~\cite{quach2018piecewise}. Tree shaking and dead-code elimination remove code that is statically unreachable from a particular build, bundle, or execution path~\cite{webpack_treeshaking, malavolta2023deadcode}. Vendoring internalizes a dependency by copying the upstream source code directly into the downstream repository, often to improve control, reproducibility, or availability~\cite{pip_vendoring_policy}. These strategies are highly useful because they explicitly recognize that the full adopted dependency often exceeds what a downstream project actually needs.

\begin{table}[t]
\caption{Comparison of dependency reduction and replacement strategies. Use-case-oriented regeneration differs because it targets the repository-used slice of functionality as the unit of sourcing and replacement.}
\label{tab:strategy-comparison}
\centering
\footnotesize 
\setlength{\tabcolsep}{4pt} 
\begin{tabular}{p{0.21\linewidth}p{0.20\linewidth}p{0.29\linewidth}p{0.22\linewidth}}
\toprule
\textbf{Approach} & \textbf{Starts from} & \textbf{Main goal} & \textbf{Relies on full upstream artifact?} \\
\toprule
Debloating & Adopted artifact & Remove unnecessary code and reduce attack surface & Yes, typically \\
Tree shaking / DCE & Adopted artifact and build graph & Remove unreachable shipped code & Yes \\
Vendoring & Upstream artifact & Internalize code for local control and patching & Yes, conceptually \\
Full regeneration \newline (\eg HARP, Lexo) & Upstream artifact & Synthesize a full replacement of the package & No \\
\midrule
\textit{Use-case-oriented regeneration} & Repository--dependency pair & Replace repository-used functionality with local impl. & Not necessarily \\
\bottomrule
\end{tabular}
\end{table}

More recently, \emph{full regeneration} strategies, such as HARP~\cite{Vasilakis2021Harp} and Lexo~\cite{lamprou2025lexo}, have emerged to eliminate supply-chain attacks by leveraging program inference and LLMs to synthesize complete, semantically equivalent replacements for upstream packages.

However, as highlighted in Table~\ref{tab:strategy-comparison}, these strategies generally begin \emph{after} a dependency has already been adopted as an upstream artifact, or they treat the \emph{upstream package} as the fundamental unit of analysis. Debloating and tree shaking reduce unused code from an existing dependency, while vendoring moves upstream source into the downstream repository but still relies on the upstream implementation as the starting point. Full regeneration attempts to replace the artifact entirely, but still aims to synthesize a general-purpose, 1-to-1 package equivalent of the third-party library.

\section{Use-Case-Oriented Regeneration}
\label{sec:regeneration}

We propose \emph{use-case-oriented regeneration} to address the structural liabilities of artifact-centric reuse. We define this approach as the local synthesis of only the specific dependency functionality that a downstream repository actually exercises, treating the \emph{repository--dependency pair} as the fundamental unit of replacement. Crucially, this targets \emph{repository-observed equivalence} rather than full semantic equivalence with the upstream package.
A generated replacement implements only the behaviors directly exercised by the repository, allowing unused API surfaces to be discarded, and success to be evaluated by whether the repository's validation artifacts continue to pass.

We can view use-case-oriented regeneration as a projection problem.
Let \(P\) denote an upstream package, and let~\(R\) denote a downstream repository that uses~\(P\).
The repository induces an observation context \(O_R\): the set of calls, inputs, configurations, validation artifacts, and execution paths through which \(R\) exercises \(P\).
Unlike full regeneration~\cite{Vasilakis2021Harp,lamprou2025lexo}, which seeks a replacement \(P'\) such that \(P' \equiv P\) over the full package semantics,
\textit{use-case-oriented regeneration} seeks a local implementation \(L_{R,P}\) such that
\[
  \pi_{O_R}(L_{R,P}) \equiv \pi_{O_R}(P),
\]
where the projection \(\pi_{O_R}\) projects package behavior onto the behaviors observable through \(R\)'s use of the dependency.
Thus, the relevant equivalence class is defined by the repository--dependency pair, not by the upstream artifact alone.

This paradigm changes the locus of supply chain trust. In the current model, developers rely on external signals---such as package identity, ecosystem provenance, and maintainer reputation---that are often decoupled from the actual code executed~\cite{Hou2023TrustSECO, Hamer2025TrustingCode, SLSAProvenance}. Regeneration reduces runtime reliance on these external signals, shifting the trust problem from evaluating an upstream ecosystem to the \emph{local verification} of synthesized code. 
The repository owner must therefore establish adequate evidence for the correctness, test coverage, and maintainability of a hyper-specialized local implementation.

\section{Study Design}
\label{sec:study-design}

This section describes the empirical study used to evaluate use-case-oriented regeneration.
We evaluate whether an agentic workflow can regenerate the functionality actually used by a client repository across a deliberately varied benchmark of repository--dependency pairs.
Our study is a proof-of-concept: each successful regeneration provides evidence that a bounded slice of dependency functionality can be replaced by validated local code, while failures expose limits of the current workflow.
Because regeneration is repository-specific and judgment-laden (cf.~\cref{sec:background:davis}), the fully automated workflow described here is illustrative rather than prescriptive.

\subsection{Research Questions and Evaluation Metrics}
\label{subsec:rqs-and-metrics}

We ask three research questions, each paired with an evaluation metric:

\begin{description}
    \item[RQ1 (Feasibility).] Given a client repository and a dependency it uses, can an agent regenerate the exercised dependency functionality to remove first-party runtime reliance? \newline
    \emph{Metric: Successful removal of the target dependency from runtime code and the successful integration of an executable local replacement.}
    
    \item[RQ2 (Behavioral Preservation).] To what extent do generated implementations preserve repository behavior across repository--dependency pairs? \newline
    \emph{Metric: The aggregate pass rate of the repository's baseline validation checks (tests, builds, linters) before and after regeneration.} 
    
    \item[RQ3 (Size and Footprint).] How do generated implementations compare with the original dependency in terms of size and dependency footprint? \newline
    \emph{Metric: The measured reduction in exported API surface and implementation footprint compared to the original upstream package.}
\end{description}

Taken together, these metrics evaluate both whether regeneration works as a repository transformation and what kind of artifact it produces when it succeeds. This distinction is critical: a generated replacement may pass repository validation while still raising concerns about untested behavior or maintainability.

\subsection{Subject Dependencies and Repositories}
\label{subsec:subjects}

We evaluate use-case-oriented regeneration on nine JavaScript/TypeScript dependencies selected to vary in size, role, popularity, and expected regeneration difficulty. The dependencies span small utility packages, medium-sized libraries, and larger packages with broader API surfaces. This selection is not intended to be statistically representative of all npm packages; instead, it is designed to exercise regeneration across a range of dependency types for which repository-specific replacement may plausibly range from straightforward to difficult~\cite{wittern2016look, zimmermann2019npm}.

\begin{table}[htbp]
\caption{Target dependencies used in the study, grouped by approximate package size.}
\label{tab:target-dependencies}
\centering
\begin{tabular}{p{0.16\linewidth}p{0.20\linewidth}p{0.46\linewidth}c}
\toprule
\textbf{Size tier} & \textbf{Dependency} & \textbf{Type / role} & \textbf{Pairs} \\
\toprule
Small & \texttt{nanoid} & Identifier generation & 20 \\
Small & \texttt{change-case} & String case conversion & 20 \\
Small & \texttt{chalk} & Terminal string styling & 20 \\
Medium & \texttt{express} & Web server framework & 20 \\
Medium & \texttt{semver} & Version parsing and range handling & 20 \\
Medium & \texttt{postcss} & CSS transformation framework & 20 \\
Large & \texttt{lodash} & General-purpose utility library & 20 \\
Large & \texttt{axios} & HTTP client & 20 \\
Large & \texttt{zod} & Schema validation & 20 \\
\midrule
\textit{Total} & \textit{9 dependencies} &  & \textit{180} \\
\bottomrule
\end{tabular}
\end{table}

For each dependency, we select 20 evaluable client repositories (180 pairs total). We identified candidate repositories through a code search over public GitHub repositories that imported the target dependencies. This initial pool was filtered to retain only repositories suitable for evaluation:
  they had to build in a clean environment, expose validation artifacts (such as test suites or executable scripts), and pass their baseline validation prior to any regeneration attempt. 
  When multiple repositories were equally suitable, repository popularity was used as a secondary preference signal.
  The primary unit of analysis remains the repository--dependency pair, as regeneration feasibility depends not only on the dependency itself but also on how deeply a particular repository uses it.

\subsection{Regeneration Procedure}
\label{subsec:regeneration-procedure}

\begin{figure}[htbp]
    \centering
    \includegraphics[width=0.95\linewidth]{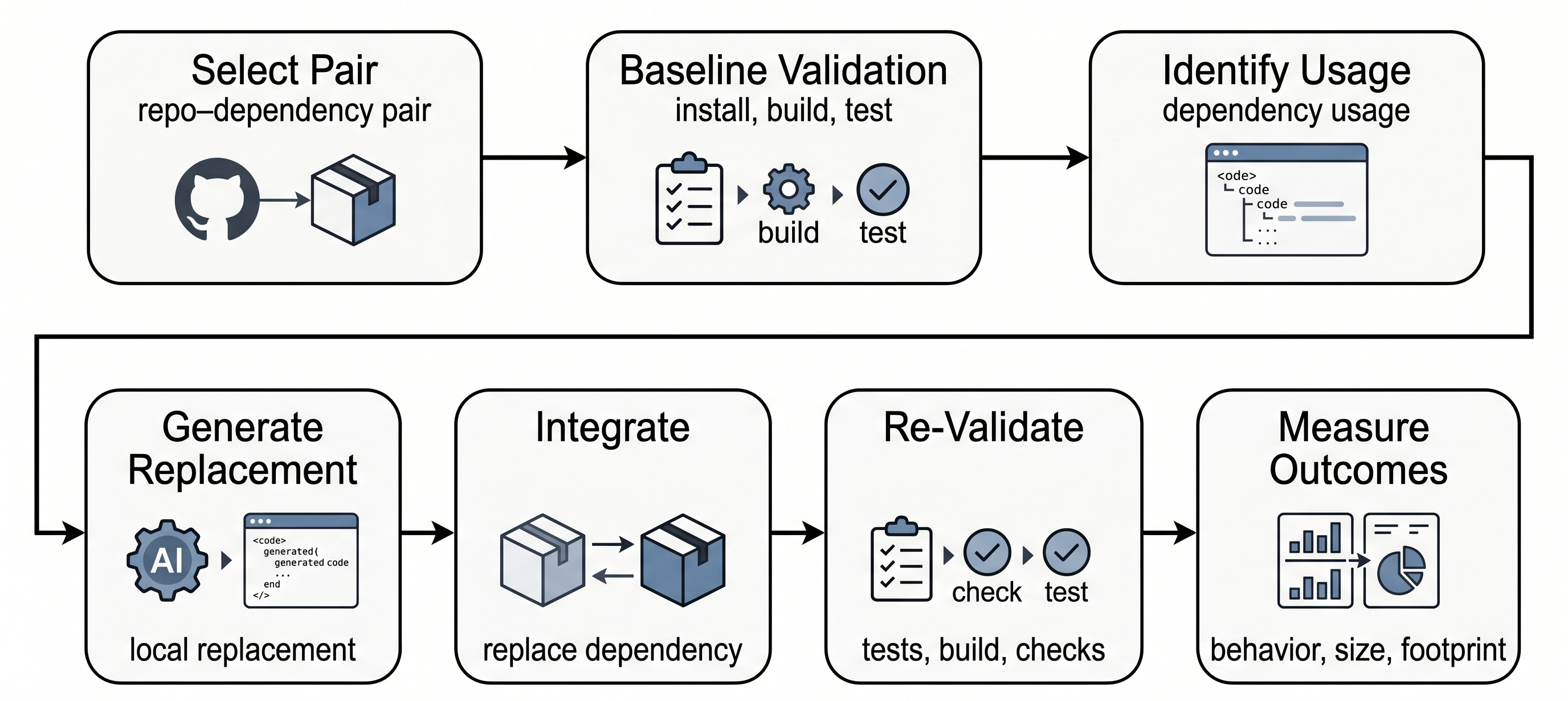}
    \caption{Our experimental workflow. For implementation, see our artifact.}
    \label{fig:workflow}
\end{figure}

The experimental workflow is summarized in \Cref{fig:workflow}. For each of the 180 pairs, we:

\begin{enumerate}
    \item \textbf{Establish Baseline:} We clone the repository at a fixed revision, install its declared dependencies, and run its validation artifacts. Only repositories with successful baseline validation are retained, providing a strict behavioral denominator for comparison.
    \item \textbf{Regenerate Agentically:} The agent is given access to the repository context and validation artifacts. Its task is to identify first-party usages of the target dependency, generate local replacement code, update call sites, remove dependency declarations, and iterate on validation failures.
    \item \textbf{Re-Validate:} The output is a modified repository. We rerun the baseline validation artifacts and measure the feasibility, behavioral preservation, and footprint differences.
\end{enumerate}

\section{Results}
\label{sec:emerging-results}

This section reports results from our evaluation of use-case-oriented regeneration across 180 repository--dependency pairs.
We interpret these results as a characterization of the feasibility and limits of repository-centered regeneration. 

\subsection{RQ1 \& RQ2: Feasibility and Behavioral Preservation}
\label{subsec:rq1-rq2-feasibility-preservation}

As detailed in \cref{tab:regeneration-summary}, use-case-oriented regeneration proved feasible and highly behavior-preserving across a substantial fraction of the evaluated pairs. In aggregate, generated replacements achieved a 99.8\% validation pass rate, with 166 of the 180 pairs demonstrating perfect behavioral preservation.
The agent identified first-party call sites, generated local replacement code, updated callers, and removed runtime reliance on the original packages. 

However, outcomes varied by dependency. While highly bounded utilities like \texttt{nanoid} and \texttt{zod} saw perfect preservation across all client repositories, more complex or deeply integrated dependencies like \texttt{lodash} presented greater challenges. This variation confirms that feasibility depends heavily on both the target dependency's complexity and how specifically the downstream repository uses it.

Crucially, this high rate of preservation must be interpreted carefully. Because preservation is measured using each repository's existing validation artifacts, these results confirm that regeneration can seamlessly preserve \emph{repository-observed} behavior. They do not prove full semantic equivalence across all possible inputs, as a generated replacement may pass available tests while omitting untested edge cases or undocumented compatibility behaviors.

\subsection{RQ3: Size and Dependency Footprint}
\label{subsec:rq3-footprint}

Successful replacements were substantially narrower than their upstream counterparts. As shown in \cref{tab:api-surface-reduction}, the generated implementations exposed drastically smaller interfaces—averaging a 93.1\% reduction in exported API surface across the benchmark. 

\begin{table}[t]
    \caption{Summary of regeneration outcomes across 180 repository--dependency pairs. ``Baseline'' and ``After'' report the number of validation checks passing before and after regeneration. ``Perfect'' reports the number of repository--dependency pairs for which all baseline checks still pass after regeneration.}
    \label{tab:regeneration-summary}
    \centering
    \begin{tabular}{p{0.18\linewidth}c r r r c}
    \toprule
    \textbf{Dependency} & \textbf{Repos} & \textbf{Baseline} & \textbf{After} & \textbf{Failed} & \textbf{Perfect} \\
    \toprule
    \texttt{axios} & 20 & 7,513 & 7,509 & 4 & 18/20 \\
    \texttt{chalk} & 20 & 4,716 & 4,715 & 1 & 19/20 \\
    \texttt{change-case} & 20 & 20,917 & 20,880 & 37 & 19/20 \\
    \texttt{express} & 20 & 1,170 & 1,164 & 6 & 18/20 \\
    \texttt{lodash} & 20 & 12,804 & 12,700 & 104 & 16/20 \\
    \texttt{nanoid} & 20 & 2,118 & 2,118 & 0 & 20/20 \\
    \texttt{postcss} & 20 & 2,292 & 2,289 & 3 & 18/20 \\
    \texttt{semver} & 20 & 5,111 & 5,109 & 2 & 18/20 \\
    \texttt{zod} & 20 & 7,993 & 7,993 & 0 & 20/20 \\
    \midrule
    \textit{Total} & \textit{180} & \textit{64,634} & \textit{64,477} & \textit{157} & \textit{166/180} \\
    \bottomrule
    \end{tabular}


    \caption{API-surface reduction in generated replacements. The generated implementations expose substantially smaller interfaces than the original dependencies, reflecting repository-specific usage rather than full-package replacement.}
    \label{tab:api-surface-reduction}
    \centering
    \begin{tabular}{p{0.22\linewidth}r r r r}
    \toprule
    \textbf{Dependency} & \textbf{Original exports} & \textbf{Avg. regenerated} & \textbf{Reduction} & \textbf{Utilization} \\
    \toprule
    \texttt{zod} & 236 & 4.1 & 98.3\% & 1.7\% \\
    \texttt{lodash} & 306 & 19.4 & 93.7\% & 6.3\% \\
    \texttt{chalk} & 64 & 5.0 & 92.2\% & 7.8\% \\
    \texttt{postcss} & 25 & 1.0 & 96.0\% & 4.0\% \\
    \texttt{change-case} & 14 & 0.4 & 97.1\% & 2.9\% \\
    \texttt{axios} & 33 & 6.4 & 80.6\% & 19.4\% \\
    \texttt{semver} & 45 & 9.8 & 78.1\% & 21.9\% \\
    \texttt{nanoid} & 5 & 1.4 & 73.0\% & 27.0\% \\
    \texttt{express} & 11 & 3.2 & 70.9\% & 29.1\% \\
    \midrule
    \textit{Overall} & \textit{82.1} & \textit{5.6} & \textit{93.1\%} & \textit{6.9\%} \\
    \bottomrule
    \end{tabular}
\end{table}

\begin{wrapfigure}[12]{r}{0.52\textwidth}
  \centering
  \includegraphics[width=\linewidth]{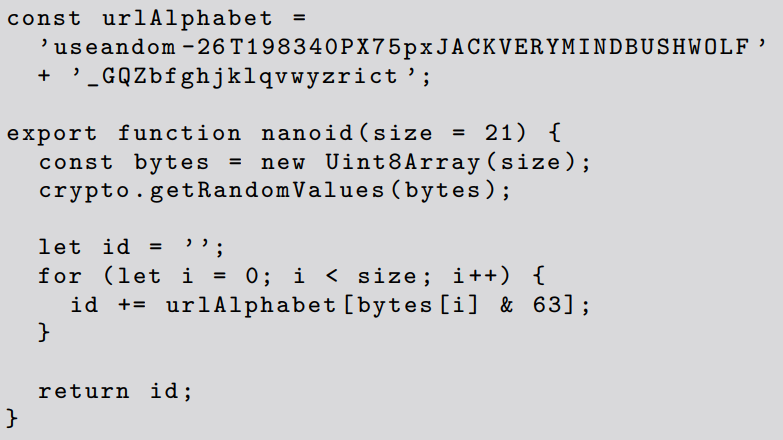} 
  \caption{Use-case-regenerated replacement for \texttt{nanoid}.}
  \label{fig:nanoid}
\end{wrapfigure}

This reduction empirically validates the core motivation for this strategy: downstream repositories typically rely on a highly bounded slice of a dependency's potential behavior.
To illustrate this footprint reduction, Figure~\ref{fig:nanoid} shows a generated replacement for \texttt{nanoid}.
The agent successfully extracted only the core random-byte generation logic required by the client, shedding the original package's unused CLI tools and custom alphabets.
This reduces the dependent's unused API surface and transitive exposure.

Footprint reduction alone is an imperfect proxy for quality.
A smaller replacement is easier to audit, but may omit behaviors that future repository evolution might require.
The primary empirical signal here is that these replacements are significantly smaller \emph{while still preserving the current validation behavior}, suggesting regeneration is most effective for narrow, stable, and well-tested usage slices.

\subsection{Failure Modes and Limitations}
\label{subsec:failure-modes}

Analyzing the 14 failed regeneration attempts clarifies the current limits of this approach. Failures typically emerged when the target dependency provided broad, subtle, or highly dynamic behaviors that were difficult for the agent to infer strictly from static call sites. Specifically, we observed three primary failure modes:
\begin{enumerate}
    \item \textit{Semantic and edge-case mismatches}, such as failing to reproduce \texttt{semver}'s prerelease logic or \texttt{lodash}'s deep cloning and chained APIs;
    \item \textit{Class identity mismatches}, where generated objects failed strict \texttt{instanceof} checks against native or cross-package prototypes (\eg in \texttt{zod} and \texttt{semver});
    and
    \item \textit{Deep framework integrations}, where replacements could not satisfy complex internal tooling expectations, such as \texttt{express} middleware arity or \texttt{axios} mock interceptors. 
\end{enumerate}
\noindent In these scenarios, the agent successfully handled the common path but failed on nuanced edge cases exercised by the repository's test suite.
These failures may also reflect the limitations of our validation setup. Repository validation artifacts can be brittle, environment-sensitive, or uneven in coverage. Passing a weak test suite might hide significant semantic deviations, while failing a flaky or snapshot-sensitive test might falsely indicate a broken replacement.

Detailed descriptions of these attempts are given in our accompanying artifact.

\section{Discussion and Research Agenda}
\label{sec:discussion}

The results indicate that use-case-oriented regeneration is a potentially viable sourcing strategy.
Many repositories rely on a significantly narrower slice of functionality than the adopted package provides, and that slice can be regenerated. 
This suggests a broader research direction: treating dependency replacement as a repository-centered sourcing problem rather than a package-level synthesis problem.

\mysubsection{When Is Regeneration Appropriate?}
Our results suggest that regeneration is most promising when a repository uses a small, well-localized, and well-tested slice of a dependency. Utility-style libraries and narrow API usage are the cases where the footprint reduction and security benefits are clearest. Conversely, regeneration is riskier when the dependency entails complex framework behavior, subtle semantics, or security-sensitive logic that is difficult to infer from static call sites.

Consequently, we suggest that use-case-oriented regeneration should be developed as an \emph{assisted software engineering workflow} rather than a fully automatic dependency removal mechanism. The most realistic near-term application is human-reviewed generation of local replacements for stable, narrow dependencies (\textit{cf.}~\cref{sec:background:davis}). By establishing a decision framework that weighs dependency role, usage breadth, and validation strength, we can transition from a model of rigid upstream adoption to one of flexible, verifiable, and repository-specific generation.

\mysubsection{Behavioral Confidence and Verification.}
The central challenge for regeneration is behavioral confidence. In our study, preservation was measured using existing repository validation artifacts. While pragmatic, passing existing tests does not guarantee safety across unexercised edge cases or undocumented compatibility behaviors. 
Under conventional reuse, weak downstream test-characterization of dependency semantics is often tolerated because the upstream artifact is the behavioral authority~\cite{winters2020software}.
Regeneration changes that assumption:
  once the dependency is replaced by local code, stronger evidence is needed for the exercised dependency semantics.
Future work must therefore establish verification workflows that go beyond existing test suites.
Directions might include differential testing, 
coverage-guided validation, and focused test generation for the repository-used API slice.

\mysubsection{Implications for Engineering Processes.}
%
Use-case-oriented regeneration changes how downstream projects benefit from the ``many eyes'' of open-source reuse~\cite{raymond1999cathedral}.
In conventional artifact-centric reuse, a widely used upstream package accumulates not only implementation fixes, but also specification knowledge: API conventions, clarified edge-case behavior, documentation, tests, examples, and compatibility expectations.
These are precisely the properties that afford regeneration.
A downstream project that continues to depend on the package can inherit this evolving specification through dependency updates.
A regenerated replacement, by contrast, captures a repository-specific slice of that specification at generation time.
It does not automatically inherit later semantic changes, security fixes, or API clarifications.
Regeneration therefore entails an explicit refresh practice: monitoring upstream changes, deciding whether they affect the local API slice, and acting when they do.
This situation demands regeneration provenance metadata. 
Without such metadata, maintainers cannot reliably determine whether a later upstream change is irrelevant, requires regeneration, or exposes an unsound local approximation.

\section{Threats to Validity}
\label{sec:threats}
As this is a vision paper with a limited measurement, several threats affect our findings: 
\begin{itemize}
    \item \textbf{Construct Validity:} Preservation is measured via existing repository tests, not formal proofs. We mitigate this by bounding claims strictly to \emph{repository-observed} equivalence.
    \item \textbf{Internal Validity:} Outcomes rely on a specific agentic workflow (Claude Opus) and prompt. We isolated agent impact by requiring strict baseline pre-validation.
    \item \textbf{External Validity:} Evaluation is limited to JS/TS and to 9 dependencies. Feasibility may differ in ecosystems with different testing cultures or agent training data.
\end{itemize}

\section{Conclusion}
\label{sec:conclusion}
Modern software systems rely on complex dependency graphs with unanticipated ongoing maintenance costs.
We argue that the traditional build-\textit{vs.}-reuse tradeoff demands reexamination.
We envision \emph{use-case-oriented regeneration} as a strategy in which generative AI locally synthesizes the necessary slice of an upstream dependency.
Our evaluation of 180 repository--dependency pairs suggests that agentic workflows can often replace adopted packages with narrow implementations that preserve repository-observed behavior.
However, realizing this vision safely requires concerted research into the maintainability and trustworthiness of generated code.
Because existing validation artifacts are imperfect safety proxies, future work must prioritize coverage-guided validation, differential testing, and AI-assisted risk assessment workflows.
Ultimately, use-case-oriented regeneration can shift the software supply chain from rigid upstream adoption to flexible, verifiable, and repository-specific generation, granting downstream projects greater control over the code they trust and maintain.

\clearpage


\section*{Data Availability}
\label{app:extended-dataset}
We provide an anonymous replication package.
It includes all prompts, dataset, replication scripts, and an extended version of the manuscript containing additional tables.
Visit it at:
  \url{https://doi.org/10.5281/zenodo.20370439}.

\vspace{-0.1cm}

\bibliography{lipics-v2021-sample-article}

\end{document}